%% file: routing-forwarding-ntorrent.tex
\documentclass[10pt, conference, letterpaper]{IEEEtran}

\usepackage{hyperref}
\usepackage{array}
\usepackage{graphicx}
\usepackage[table]{xcolor}
\usepackage{algpseudocode}
\usepackage{hyperref}

\newcolumntype{C}[1]{>{\centering\let\newline\\\arraybackslash\hspace{0pt}}m{#1}}
\algnewcommand\algorithmicinput{\textbf{Input:}}
\algnewcommand\INPUT{\item[\algorithmicinput]}



\title{Routing and Forwarding in nTorrent using ndnSIM}
\author{
\IEEEauthorblockN{Akshay Raman}
\IEEEauthorblockA{University of California, Los Angeles \\  akshay.raman@cs.ucla.edu}
}

\usepackage{etoolbox}
\usepackage{xstring}
\DeclareListParser{\doslashlist}{/}
\newcounter{ndnNameComponentCounter}%
\newcommand{\ndnName}[1]{{%
  \setcounter{ndnNameComponentCounter}{0}%
  \renewcommand{\do}[1]{{%
    \ifnumgreater{\value{ndnNameComponentCounter}}{0}{\allowbreak/}{}%
    \ifnumodd{\value{ndnNameComponentCounter}}{}{}%
    \detokenize{##1}}%
    \stepcounter{ndnNameComponentCounter}}%
``{\fontfamily{cmtt}\small\selectfont\IfBeginWith{#1}{/}{/}{}\doslashlist{#1}}''%
}}

\usepackage{eso-pic,xcolor}
\makeatletter
\AddToShipoutPicture*{%
\setlength{\@tempdimb}{20pt}%
\setlength{\@tempdimc}{\paperheight}%
\setlength{\unitlength}{1pt}%
}
\makeatother

\usepackage{color,soul}
\usepackage{xspace}

\begin{document}
\maketitle

\begin{abstract}

BitTorrent is a popular communication protocol for peer-to-peer file sharing. It uses a data-centric approach, wherein the data is decentralized and peers request each other for pieces of the file(s). Aspects of this process is similar to the Named Data Networking (NDN) architecture, but is realized completely at the application level on top of TCP/IP networking. nTorrent is a peer-to-peer file sharing application that is based on NDN. The goal of this project is to port the application onto ndnSIM and illustrate the effects of routing and forwarding.

\end{abstract}

\input{intro}
\input{related-work}
\input{design}
\input{simulations}

\input{conclusion}

\section*{Acknowledgment}

I would like to thank Spyridon Mastorakis for mentoring me during this project.

\bibliographystyle{plain}
\bibliography{refs}

\end{document}

%% file: intro.tex
\section{Introduction}

Named Data Networking (NDN)~\cite{zhang2010named} is a network layer protocol that is being actively researched with the hope of serving as a replacement for the IP protocol. nTorrent~\cite{mastorakis2017ntorrent} is an NDN peer-to-peer file sharing application. The current implementation runs with a few modifications to the base ns-3 network simulator in order to compile and run successfully. The idea behind this paper is to extend the functionality of nTorrent and make it run on top of ndnSIM~\cite{mastorakis2017evolution, mastorakis2016ndnsim, mastorakis2015ndnsim} that features full integration with the NDN Forwarding Daemon (NFD)~\cite{nfd-dev} to simulate routing and forwarding. This paper builds upon work done earlier this year~\cite{ntorrent-ndnsim-port} and includes relevant sections of that paper. For consistency, this paper uses the same topologies as that in the nTorrent~\cite{mastorakis2017ntorrent} paper.

The code is available at \url{https://github.com/akshayraman/scenario-ntorrent}.

%% file: related-work.tex
\begin{figure}[!ht]
  \centering
  \includegraphics[width=\columnwidth]{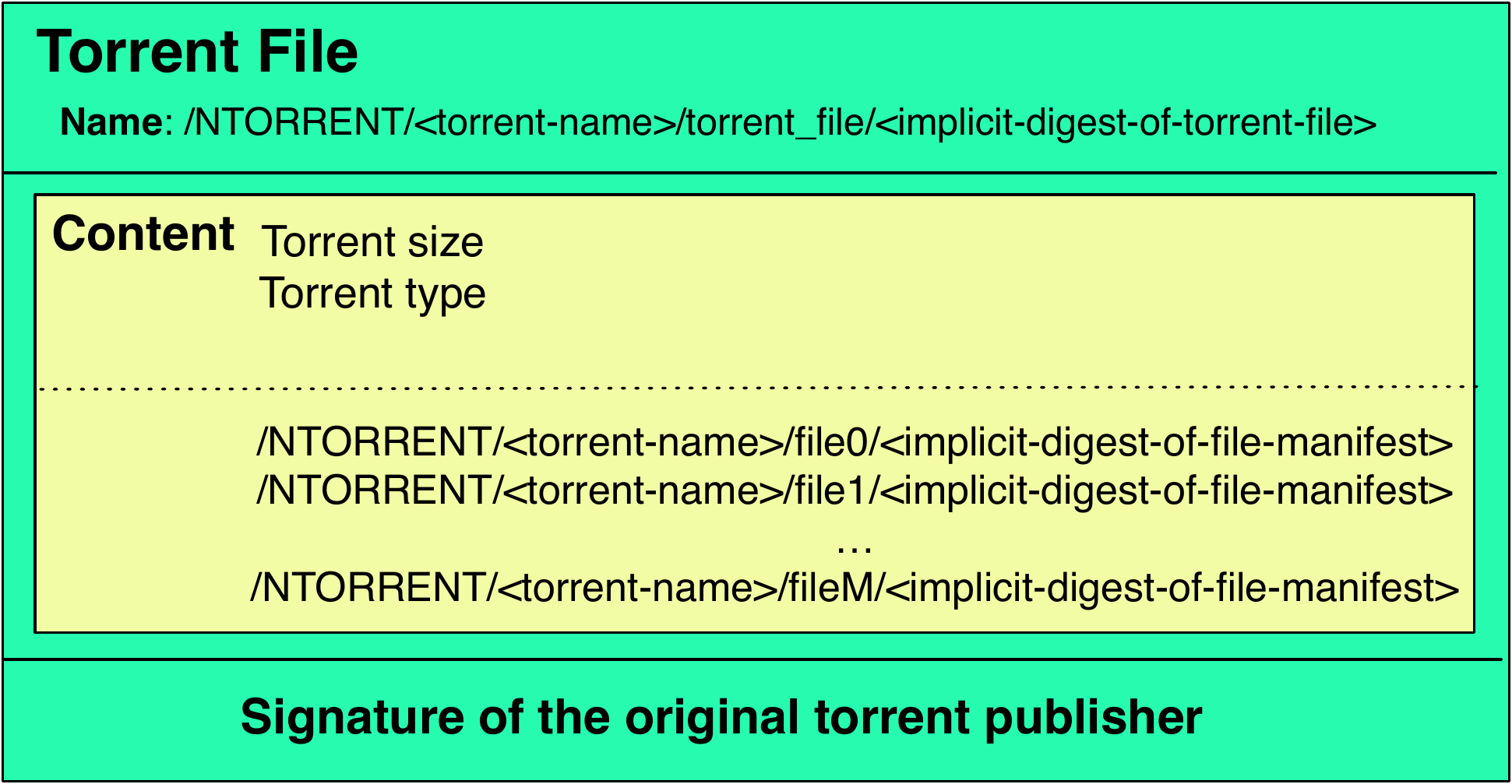}
  \caption{\small Structure of a torrent-file}
  \label{Figure:torrent-file}
\end{figure}
\begin{figure}[!ht]
  \centering
  \includegraphics[width=\columnwidth]{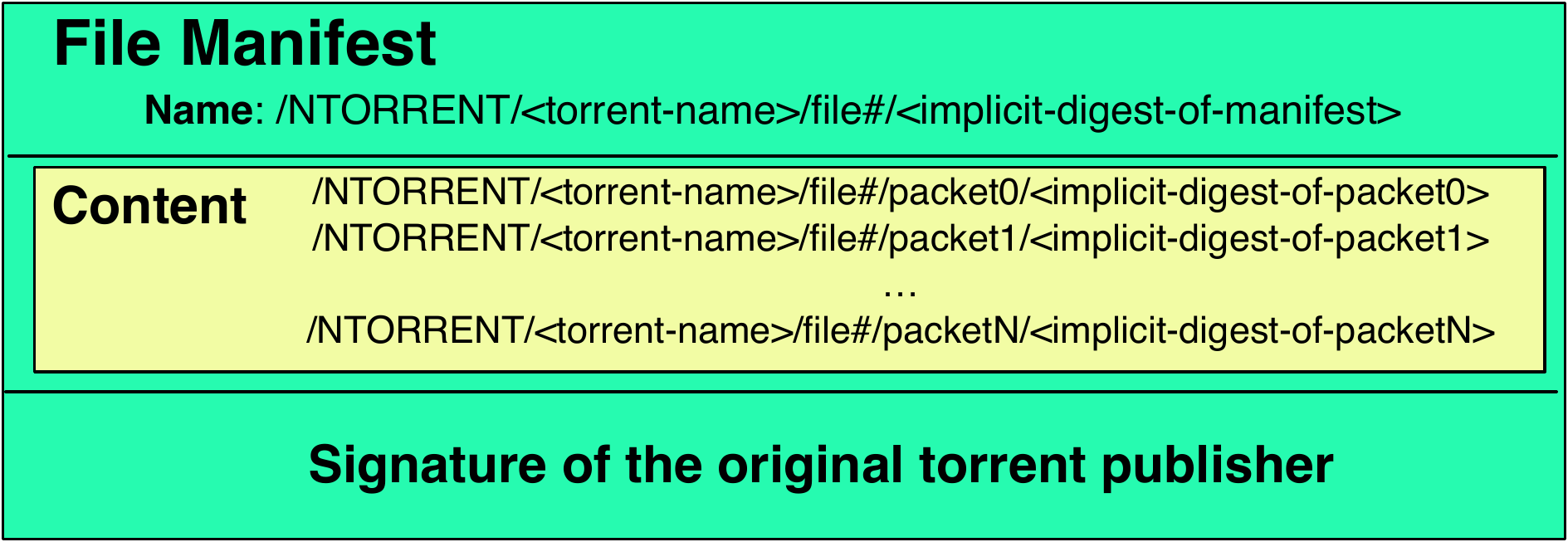}
  \caption{\small Structure of a file manifest}
  \label{Figure:file-manifest}
\end{figure}

\section{Related Work}

nTorrent has been designed to have a hierarchical file structure. At the top of the hierarchy, the .torrent file (Figure~\ref{Figure:torrent-file}) contains the name, size, type of the torrent file, and the signature of the original publisher. It also includes the names of the file manifests that make up the torrent file. Each file manifest (Figure~\ref{Figure:file-manifest}) contains its name, signature of the original publisher, and a list of names of the packets that make up that particular file. Using these names, Interest packets can be sent out to request for the corresponding files or packets. 

The fetching strategy currently implemented starts with requesting for the first packet of the first file to the last packet of the last file. Each name also follows a certain naming convention that can help easily identify the name of the torrent file, the file names, and the individual packets in a file. This name is also used by the routers to verify the integrity of the the packet.

%% file: design.tex
\section {Implementation Design}

This section describes the high-level implementation design logic and elaborates on the implementation-specific details.

\subsection {Design Logic}

ndnSIM currently supports the simulation of: (i) applications written against the ndn-cxx library (called \emph{real-world} applications) ported to ndnSIM, such as RoundSync~\cite{de2017design, roundsync-template} for decentralized data synchronization and ndn-ping~\cite{ndn-tools}, and (ii) ndnSIM-specific applications realized based on ns-3's Application abstraction, such as those developed for experimentation with Fuzzy Interest Forwarding (FIF)~\cite{Chan:2017:FIF:3154970.3154975, mastorakis2018experimentation, fif-repo} and a best-effort scheme for link reliability~\cite{vusirikala2016hop, reliability-repo}. 

The existing nTorrent code~\cite{ntorrent-code} is an application written against the ndn-cxx library, but for now, a hybrid approach between (i) and (ii) is used. Specifically, nTorrent is compiled as a shared library and import, and its data structures (e.g., torrent-file, file manifest, etc.) are used as part of the ndnSIM-specific applications. We are planning to move toward approach (i), as we further develop the codebase.

\subsection {Producer and Consumer model}

Using the ndnSIM scenario template~\cite{scenario-template} and the ChronoSync simulation example~\cite{scenario-chronosync}, we have created two extensions - one for the consumer and the other for the producer. The most basic simulation in this project is called "ntorrent-simple" and this is the entry point for the simulation. In this script, we configure the attributes of the system (data rate, latency and so on) as well as the individual attributes for the producer and consumer. These attributes include number of names per torrent segment, number of names per manifest and the size of each data packet. All these variables have default values but can be easily modified via the command-line.

The producer is responsible for generating the torrent file segments, manifests and data packets. The code to generate the above is being reused from the existing nTorrent implementation. All of this is stored in private variables belonging to the producer. The number of files being transferred is a constant that can be configured in the simulation-constants.hpp file. These are not real files - they are just pretend files to illustrate the working of nTorrent. 

The internal functionality of some modules in the nTorrent codebase was re-implemented to skip file I/O operations and use data structures (vectors) instead. For instance, all bytes read as a result of a file read operation were replaced with a dummy character "A" to simulate a file read. The producer extension also has functionality to implement the OnInterest method. This method is responsible for processing the interests that it receives from the consumer. In this method, we determine the type of the interest received. The interest could be of type torrent segment, file manifest or data packet. After determining the type of interest, the producer responds to these interests accordingly. If the interest is for a torrent file segment, the producer first checks if it has the requested segment. If so, it responds with the segment. The file manifest and data packet interest-response functionality is also handled similarly. 

The consumer has a copyTorrentFile method to simulate the copy of the torrent file. In the BitTorrent model, the consumer and producer are both expected to have the torrent file. This is achieved in the simulation by generating the torrent file on the consumer end too. As the NDN communication model is consumer-driven, the consumer starts the file transfer with an interest for the first torrent segment, contained in this torrent file. Additionally, the consumer implements the OnData and SendInterest methods. The SendInterest method is used to send out an interest packet. This method appends a random number, the nonce, to prevent replay attacks. All interests are made using the nonce and the sha256digest. The OnData method implements the core functionality of the consumer. It is structurally similar to the OnInterest method in the producer. 

If the consumer receives a torrent file segment as the data, it uses the pointer to the next torrent segment and sends out an interest for this segment. It also iterates over the manifest catalog and sends interests for all the associated manifests. Similarly, if the consumer receives file manifest data, it uses the next manifest pointer to send an interest for the next file manifest. The associated sub-manifest catalog is iterated over to send interests for data packets. Finally, if the consumer receives a data packet, it is decrypted and displayed on the terminal. The consumer also stores all torrent segments, file manifests and data packets as private data members in vectors in lieu of actual files. The simulation terminates once all data packets have been received. The success of the file transfer can be verified using the Content Store and interface statistics. 

\subsection {Bifunctionality and Routing Announcements}

The predecessor of this paper~\cite{ntorrent-ndnsim-port} used one producer node and one consumer node. It assumed that every node acts either as a producer or a consumer. In this paper, a consumer is able to serve as a producer for a torrent-file, file manifest or data packet after it receives it from a producer (or another consumer). This is done by implementing the OnInterest method in the consumers. To simulate BitTorrent behavior, vectors are used to store the data instead of real files. Every simulation scenario features one producer that has all the data in the beginning. This is analogous to a "seeder" in BitTorrent. All other consumers start downloading the file from this producer by sending out interests and processing them as described earlier. This is analogous to "leechers" in BitTorrent. As the consumers start downloading data, they are able to serve this data to other consumers as "peers". From an NDN perspective, this involves adding a new Forwarding Information Base (FIB) entry with the name of the torrent-file, file manifest or data packet that it can serve. The FIB entry is added using the AddRoute member function of the FibHelper class. This is followed by a Routing Announcement for the name, made by the peer node. This is achieved using the AddOrigin member function of the GlobalRoutingHelper class. By adding a new origin, a peer basically tells the network that it can act as a producer for that name. Once this is done, the best path is computed for each node using the CalculateRoutes member function of the GlobalRoutingHelper class.

\subsection {Forwarding Strategy}

A good forwarding strategy will help distribute the node across the network and prevent some peers from satisfying too many interests and becoming overloaded. This can be achieved by keeping track of all incoming/outgoing interests and data at every node. If a peer can satisfy an interest sent by another peer, it responds with the data. If it cannot, it responds with a NACK. The goal of the forwarding strategy here is to ensure that the number of NACKs is minimized. The time between a node sending out an interest and receiving data should be minimal too. Taking these factors into account, there are two new metrics introduced in this paper: InterestSatisfactionRate and AverageDelay. InterestSatisfactionRate is the ratio of the number of satisfied interests to the total number of interests sent. AverageDelay is the average of all time delays between sending out an interest and receiving the data at a node.

The forwarding strategy implemented in this paper is called "ntorrent-strategy". This strategy is defined by creating a sub-class of the Strategy class from the NFD module. In the ntorrent-strategy class, the virtual callback methods afterReceiveInterest and beforeSatisfyInterest are implemented. In this strategy, the forwarding path is naively picked at first since there is no evidence of the above metrics. Once there is enough evidence of these metrics, we can use this knowledge to compute the next hop on the forwarding path. 

The afterReceiveInterest method keeps track of all incoming interest names and incoming corresponding to a particular face. This is implemented using an unordered map. The beforeSatisfyInterest method keeps track of all data packets being transmitted. When a data packet corresponding to an interest name is received, we use the timestamp logged in the afterReceiveInterest and the current timestamp to compute the delay. The arrival timestamp entry is then deleted after computing the delay. The delay value corresponding to every face is averaged out and the average delay value is used to improve the forwarding path. This is done by sorting the next hop list in increasing order of average delay.

\subsection {Code Refactoring}

To aid in code development, a lot of the code was modularized. The code in the scenario templates was cleaned up. The utility functions called createLink and createAndInstall help eliminate a lot of repetitive code. These utility functions are defined in the file "simulation-common.hpp" and can be included into any simulation scenario. Some of the simulation specific code was decoupled from the core nTorrent module. Code comments were added and documentation was improved since the predecessor~\cite{ntorrent-ndnsim-port} of this paper was written.

Though this paper talks about the 3 node degree router (ntorrent-router-node-degree-3.cpp) and 4 node degree router (ntorrent-router-node-degree-4.cpp) topologies, there are additional topologies added to the codebase. The scenarios defined in "ntorrent-multi-consumer.cpp" and "ntorrent-forwarding-scenario.cpp" were designed to test bifunctionality and routing announcements. The scenario in "ntorrent-fully-connected-consumer.cpp" is another toy example that builds upon "ntorrent-simple".

%% file: simulations.tex
\section {Simulation and Results}

\subsection {Topology}

\begin{figure}[!ht]
  \centering
  \includegraphics[width=\columnwidth]{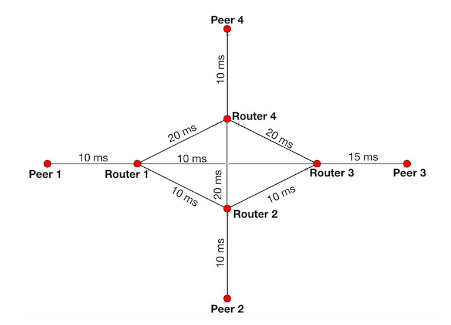}
  \caption{\small Topology with router node degree equal to 4.}
  \label{Figure:topology_routernodedeg4}
\end{figure}

\begin{figure}[!ht]
  \centering
  \includegraphics[width=\columnwidth]{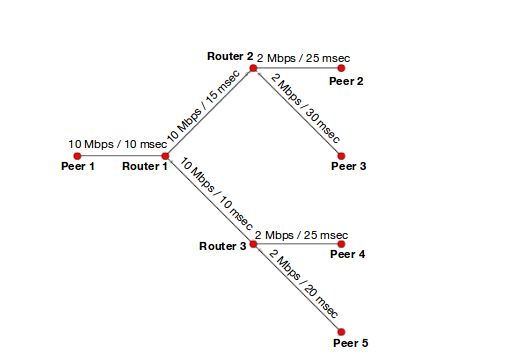}
  \caption{\small Topology with router node degree equal to 3.}
  \label{Figure:topology_routernodedeg3}
\end{figure}

To evaluate the new forwarding strategy, this paper uses the same network topologies as those described in the ~\cite{mastorakis2017ntorrent} paper. This paper deals with the topology with router node degree equal to 4 (Figure~\ref{Figure:topology_routernodedeg4}) and the topology with router node degree equal to 3 (Figure~\ref{Figure:topology_routernodedeg3}). 

The topology in Figure~\ref{Figure:topology_routernodedeg4} offers multiple paths with different costs (in terms of delays) to peers (the bandwidth of the links is the same). The one in Figure~\ref{Figure:topology_routernodedeg3} has bottleneck links to the majority of the peers. We assume that the .torrent file is known by all the peers, and the torrent to be downloaded contains 1KB data and consists of 2 files. The file sizes used in this paper are much smaller than that used in the paper~\cite{mastorakis2017ntorrent} to speed up simulation and aid in rapid design and test iterations. As a result, the bandwidth rates are modified proportionally.

In the topology described in Figure~\ref{Figure:topology_routernodedeg4}, Peer 4 acts as the seeder and announces prefixes across the network. The announcements propagate among the routers due to the routing protocol. Peer 1 starts downloading data first. After 5 seconds, peer 3 starts downloading data. After 5 more seconds, peer 2 starts its downloading; both peers 1 and 3 still act as leechers allowing us to study the effect of NDN forwarding. In the topology described in Figure~\ref{Figure:topology_routernodedeg3}, peer 1 is the seeder while peers 2, 3, 4 and 5 act as leechers. 

\begin{figure}[!ht]
  \centering
  \includegraphics[width=\columnwidth]{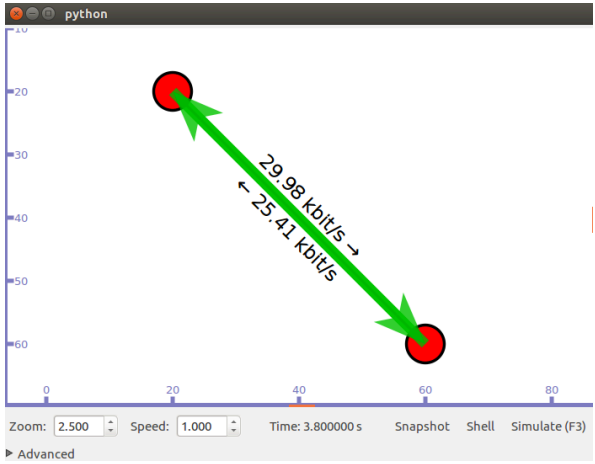}
  \caption{\small The simulation setup for "ntorrent-simple". The node in the top-left is the producer and the bottom-right node is the consumer.}
  \label{Figure:setup}
\end{figure}

This section presents a few simulation screenshots using the visualizer tool of ns-3. For the sake of simplicity, we first assume a toy example topology consisting of a node that acts as a producer and a node that acts as a consumer (Figure~\ref{Figure:setup}). The consumer node first requests the torrent-file from the producer. For now, the implemented fetching strategy\footnote{A fetching strategy that maximizes the efficiency of the retrieval process is future work.} requires that a consumer fetches all the segments of the torrent-file before it starts downloading the file manifests.

\begin{figure}[!ht]
  \centering
  \includegraphics[width=\columnwidth]{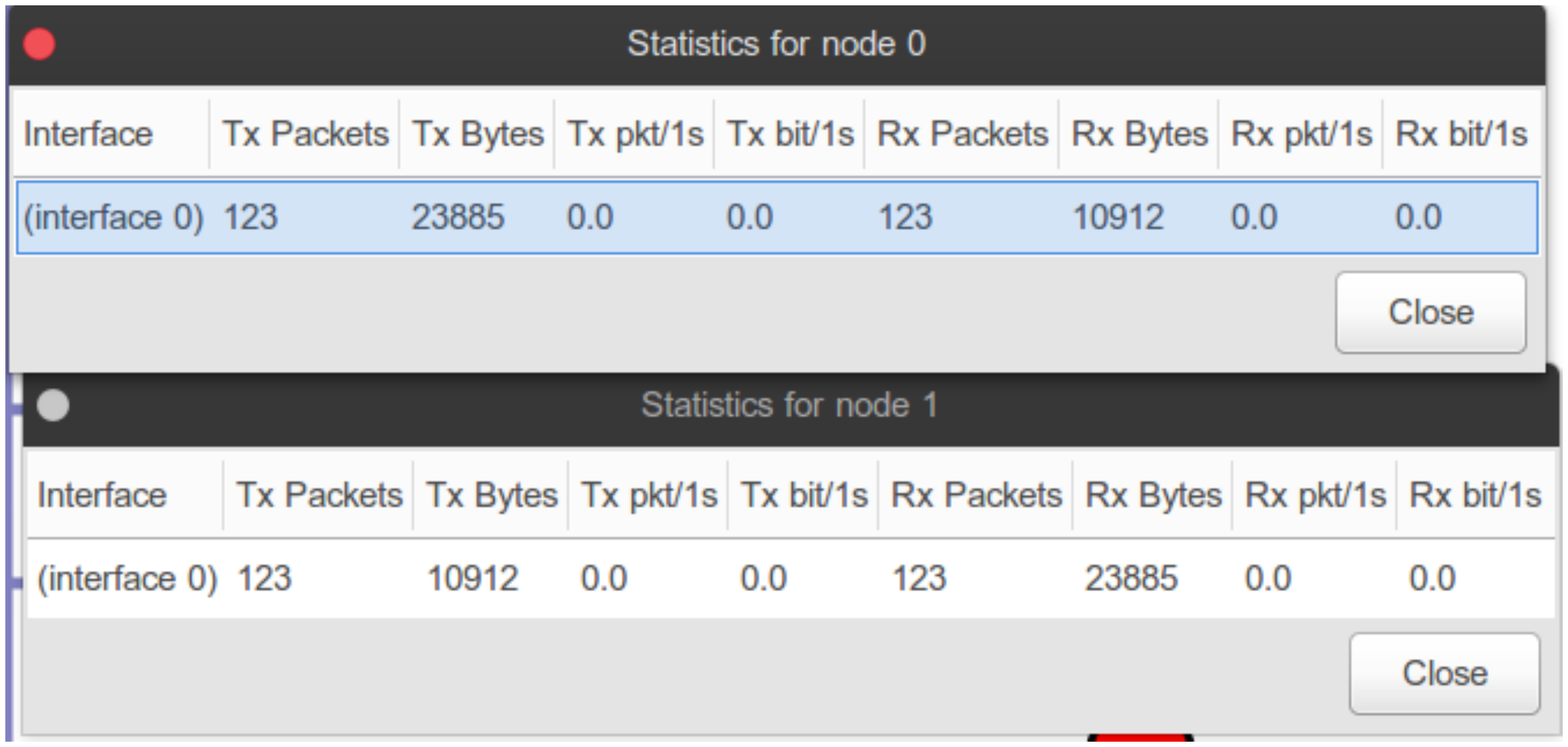}
  \caption{\small Data sent by the producer (node 0) and received by the consumer (node 1)}
  \label{Figure:node0-node1}
\end{figure}

\begin{figure}[!ht]
  \centering
  \includegraphics[width=\columnwidth]{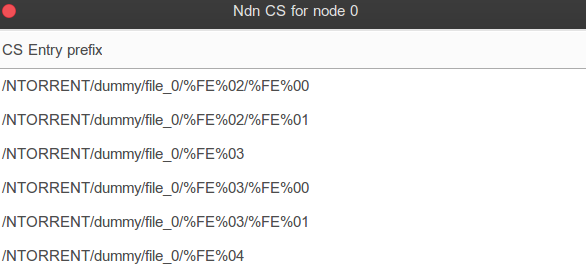}
  \caption{\small A view of the consumer's Content Store (CS)}
  \label{Figure:cs}
\end{figure}

\begin{figure}[!ht]
  \centering
  \includegraphics[width=\columnwidth]{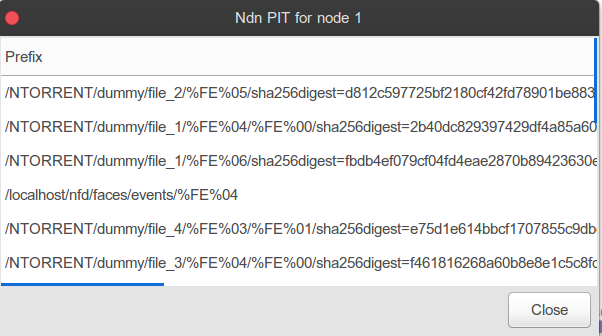}
  \caption{\small A view of the consumer's Pending Interest Table (PIT)}
  \label{Figure:pit}
\end{figure}

After all the file manifests have been fetched, the consumer starts downloading the individual data packets from the producer. Specifically, the consumer expresses Interests iteratively for all the data packets included in each file manifest. The overall process is illustrated in Figure~\ref{Figure:node0-node1} and screenshots of the consumer's CS and PIT are illustrated in Figure~\ref{Figure:cs} and~\ref{Figure:pit}.

\begin{figure}[!ht]
  \centering
  \includegraphics[width=\columnwidth]{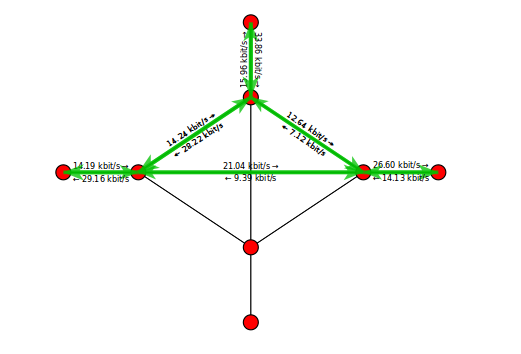}
  \caption{\small The simulation setup for "ntorrent-router-node-degree-4".}
  \label{Figure:simulation_routernodedeg4}
\end{figure}

\begin{figure}[!ht]
  \centering
  \includegraphics[width=\columnwidth]{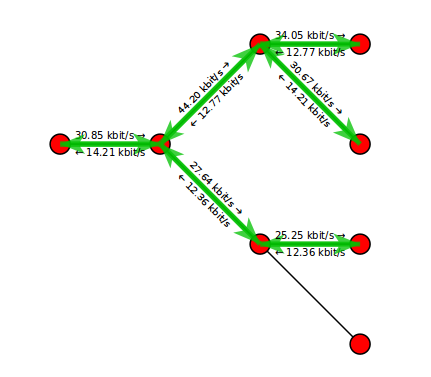}
  \caption{\small The simulation setup for "ntorrent-router-node-degree-3".}
  \label{Figure:simulation_routernodedeg3}
\end{figure}

\begin{figure}[!ht]
  \centering
  \includegraphics[width=\columnwidth]{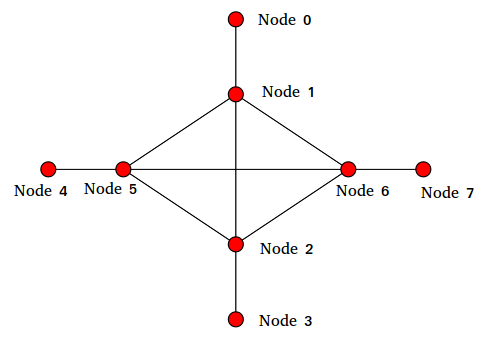}
  \caption{\small Node numbering as shown in the results graph.}
  \label{Figure:numbered_4deg}
\end{figure}

\begin{figure}[!ht]
  \centering
  \includegraphics[width=\columnwidth]{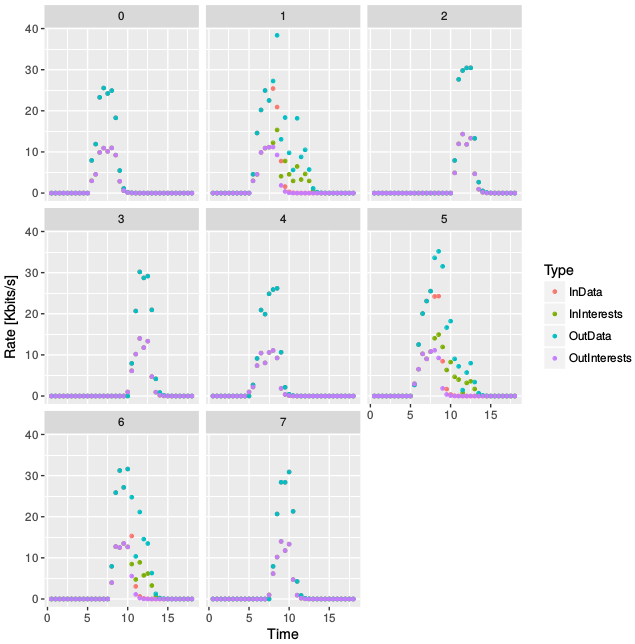}
  \caption{\small Simulation results for "ntorrent-strategy".}
  \label{Figure:results_ntorrentstrategy}
\end{figure}

\begin{figure}[!ht]
  \centering
  \includegraphics[width=\columnwidth]{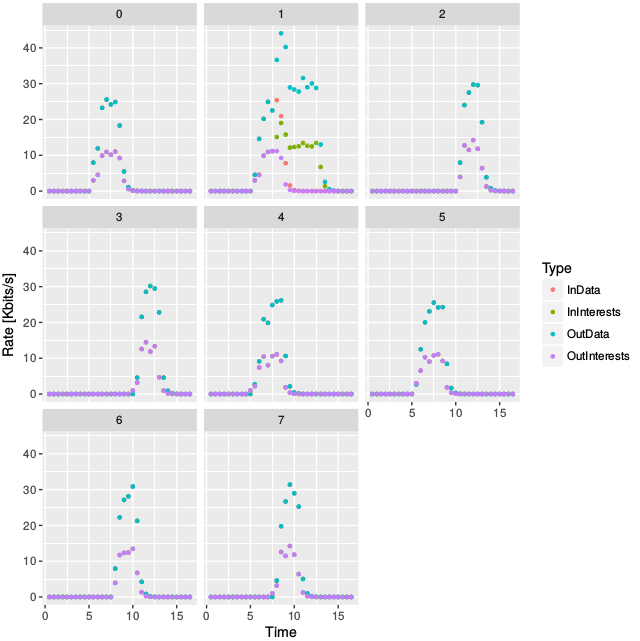}
  \caption{\small Simulation results for "client-control-strategy".}
  \label{Figure:results_cc}
\end{figure}

To better understand the results, the simulations used the L3RateTracer packet tracer to log the incoming and outgoing interests and data. This resulting log file is processed by an R script to generate the graphs in this section. The x-axis of the graph denotes the simulation time while the y-axis refers the data rate. The numbers on top of each grid in the graphs in Figures~\ref{Figure:results_ntorrentstrategy} and~\ref{Figure:results_cc} refer to the node numbers as shown in Figure~\ref{Figure:numbered_4deg}. The results in this section correspond to the topology in Figure~\ref{Figure:topology_routernodedeg4}.

On comparing the results in Figures~\ref{Figure:results_ntorrentstrategy} and~\ref{Figure:results_cc}, it is clear that the data and interests passing through nodes 5 and 6 is a higher in the ntorrent-strategy. Nodes 5 and 6 correspond to the forwarding routers 1 and 3 respectively in the Figure~\ref{Figure:topology_routernodedeg4}. The forwarding strategy ntorrent-strategy makes better use of these routers compared to the client-control strategy. The client-control strategy uses only one route to forward interests or data, whereas the ntorrent-strategy uses more than one path, if possible, as shown in ~\ref{Figure:simulation_routernodedeg4}. 

%% file: conclusion.tex
\section {Conclusion and Future Work}

This paper describes our design and experimentation with ndnSIM and nTorrent. This simulation is a clear example of how the NDN model is "consumer-driven". The ntorrent-strategy implemented in this paper performs better than the default forwarding strategies present in the ndnSIM library. This was verified by running a network-layer (L3) packet tracing logger. The results were found to be along the expected lines when compared to the findings in the paper~\cite{mastorakis2017ntorrent}. The ndnSIM mailing list and open-source communities have shown interest in this project despite it being just over a couple of months old.

The current implementation does not deal with file I/O. The plan is to support real file handling and splitting in the future. Supporting two modes of compilation (one as a standalone application and the other as an ndnSIM scenario) is another future task. Designing better fetching strategies is another open-problem that can be addressed in the future. Furthermore, the scalability and trade-offs of this nTorrent design with more tests and simulations must be investigated, especially on larger topologies.